# Ferromagnetic Ising Spin Chains Emerging from the Spin Ice under Magnetic Field

Zenji HIROI, Kazuyuki MATSUHIRA[1] and Masao OGATA[2]

*Institute for Solid State Physics, University of Tokyo, Kashiwa, Chiba 277-8581*
[1]*Department of Electronics, Faculty of Engineering, Kyushu Institute of Technology, Kitakyushu 804-8550*
[2]*Department of physics, University of Tokyo, Hongo, Tokyo 113-0033*

A spin-ice compound dysprosium titanate, $Dy_2Ti_2O_7$, is studied by specific heat measurements in magnetic fields applied along the [110] direction of the cubic unit cell. Above a magnetic field of 0.4 T a relatively sharp peak at $T = 1.1$ K in the specific heat appears from a broad peak associated with spin-ice freezing at low fields. This new peak is almost independent of field up to 1.5 T, while another broad, field-dependent peak is detected at higher temperature. Under magnetic field along [110] Ising spins on the pyrochlore lattice can be considered to form two orthogonal sets of chains; parallel and perpendicular to the field direction. It is suggested that these perpendicular chains behave as unique ferromagnetic Ising spin chains without long-range order, giving rise to the observed field-independent peak.

Geometrical frustration between localized spins on some symmetrical lattices causes often interesting physics. The spin ice is a typical example in which a large Ising spin from a rare-earth ion resides on the pyrochlore lattice made of corner-sharing tetrahedra and weakly interacting ferromagnetically with its neighbors.[1,2] The local quantization axis is fixed to the <111> directions of the cubic unit cell due to the crystal field effect. Consequently, a two-in, two-out spin configuration in each tetrahedron is preferred, which is called the ice rule. However, since there still remains a macroscopic number of ways to span the whole pyrochlore lattice, macroscopic degeneracy in the ground state analogous to water ice is realized.[3] This peculiar ground state with residual entropy of about 1.7 J/K mol Dy was found experimentally in three pyrochlore oxides, $Ho_2Ti_2O_7$,[1] $Dy_2Ti_2O_7$,[4] and $Ho_2Sn_2O_7$.[5] It has been pointed out that the formation of the spin ice state in these real materials is due to long-range dipole-diploe interactions,[6] which has been evidenced by measuring the zero-field spin correlations in elestic neutron-scattering experiments on a $Ho_2Ti_2O_7$ single crystal.[7,8]

Recently, Matsuhira *et al.*[9,10] reported another macroscopically degenerate state in $Dy_2Ti_2O_7$ in a magnetic field applied along the [111] direction. This new ground state is called kagomé ice,[9-11] because the degeneracy occurs in the two-dimensional kagomé net instead of the three-dimensional pyrochlore network. The pyrochlore lattice consists of triangular and kagomé planes stacked alternatively along the [111] direction. Since the spins on the former planes possess their easy axis along the field direction, they are pinned easily along the field to gain Zeeman energy. Even in this situation, however, the ice rule stabilizing the two-in, two-out configuration can be satisfied in a low magnetic field. As a result, a reduced degeneracy is expected for the kagomé plane, which is determined to be 0.67 J/K mol Dy both experimentally and theoretically.[10,12] Thus, the application of magnetic field successfully controls the dimensionality of the spin system in the spin-ice compound.

On the other hand, a field effect in the [110] direction was investigated by Fennell *et al.* based on the neutron diffraction experiments.[13] They applied a field on $Dy_2Ti_2O_7$ along the [110] direction and observed the coexistence of long-range ferrromagnetic and short-range antiferromagnetic order. This was attributed to the pinning of only half the spins by the field.[13] Moreover, magnetization measurements on $Dy_2Ti_2O_7$ showed a saturation moment of 4.08 $\mu_B$ per Dy atom, which also corresponds to that for half the spins.[14] Figure 1 illustrates this situation schematically, where the pyrochlore lattice is divided hypothetically into two sets of chains, α and β chains. A magnetic field along [110] would easily align the spins on the α chain owing to large Zeeman energy, while the β chain would remain intact. However, due to the ice rule, the spins on the β chain should also become ferromagnetically aligned at low temperature below $J/k_B$, where $J$ is an effective ferromagnetic coupling between near neighbor spins and $k_B$ is Boltzmann's constant. Note that there are still two possibilities for each $\beta$ spin chain to point right or left, though this degeneracy is not macroscopically large.

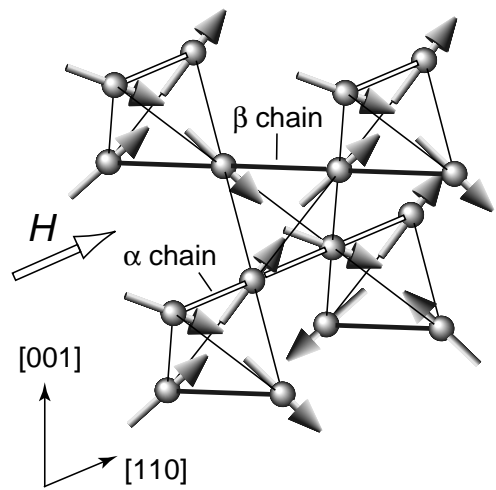

Fig. 1. Pyrochlore lattice consisting of corner-sharing tetrahedra. Balls represent Dy atoms in $Dy_2Ti_2O_7$, and arrows indicate the direction of spins that is along <111>. Under magnetic field applied along the [110] direction, the pyrochlore lattice is considered to consist of two orthogonal sets of chains, α and β chains. It is shown in the present study that this β spin chain behaves as a unique one-dimensional Ising spin system.

Here we report the specific heat of $Dy_2Ti_2O_7$ in various magnetic fields applied along the [110] direction. The purpose of the present paper is to give a thermodynamic evidence for the above-mentioned, field-induced reduction of the spin dimensionality in the spin ice. The specific heat of the $\beta$ chain is revealed and approximated theoretically by assuming a ferromagnetic Ising spin chain with long-range dipole interactions.

A single crystal of $Dy_2Ti_2O_7$ was prepared by the floating-zone method as described previously.[9] A small, thin plate-like crystal of the size $0.96 \times 1.38 \times 0.11$ mm$^3$ and the weight 0.982 mg was cut from the boule for specific heat measurements. To minimize a demagnetizing field effect, the plate surface was chosen to be (110), so that a magnetic field along [110] was parallel to the plate surface. A correction of applied magnetic field due to that effect was not taken into account in the present study.

Specific heat was measured by the heat-relaxation method in a temperature range between 0.4 K and 20 K in a Quantum Design Physical-Property-Measurement-System equipped with a $^3$He refrigerator. A single crystalline sample was attached to a sapphire substrate by a small amount of Apiezon N grease. An addenda heat capacity had been measured in a separate run without the sample at various fields, and was subtracted from the data. A magnetic field was applied approximately along the [110] direction of the cubic unit cell. The accuracy of the orientation may be within a few degrees.

Figure 2 shows the temperature dependence of magnetic specific heat $C_m$ per one mol of Dy atoms measured at various magnetic fields along the [110] direction. The $C_m$ was obtained by subtracting a lattice contribution from the raw data, as described previously.[10] At zero field it exhibits a broad maximum around $T_{p1} = 1.2$ K, in good agreement with previous results.[3] The peak temperature is close to the effective ferromagnetic interaction $J/k_B$, which has been estimated to be 1.1 K,[2] and thus freezing into the spin ice state occurs below $T_{p1}$. As magnetic field increases, first the broad peak grows upward until $\mu_0 H = 0.3$ T, and then a tiny anomaly at $T_{p2} \sim 1.0$ K becomes discernible at $\mu_0 H = 0.4$ T. With further increasing field up to 1.5 T, the anomaly turns into a relatively sharp peak at almost the same temperature. Since this field-independent peak at $T_{p2}$ is still broad, compared with that expected for long-range order (LRO), it must indicate the development of another short-range order (SRO) different from the spin ice. On the other hand, a shoulder appears above $\mu_0 H = 0.5$ T, shifts to a high-temperature side with increasing field, and finally forms another broad peak at $T_{p3}$; e.g. $T_{p3} = 7$ K at $\mu_0 H = 1.5$ T. This field dependence found for $H // [110]$ is quite different from those found for $H // [111]$ or [001].[10] The field-independent peak at $T_{p2}$ may correspond to the anomaly reported by Ramirez et al. at 1.2 K in their specific heat data on a polycrystalline sample for applied fields larger than 0.75 T.[3]

Figure 3 shows the field dependence of residual entropy $S_R$ estimated by integrating $C_m/T$ from 0.4 K or 0.5 K to 20 K, as reported previously.[10] The $S_R$ at zero field, which corresponds to the residual entropy of the spin ice, is 1.66 JK$^{-1}$mol$^{-1}$, close to the Pauling's prediction, 1.68 JK$^{-1}$mol$^{-1}$. Because of experimental difficulty in measuring $C_m$ at high temperature in high field, there is always an inevitable overestimation for $S_R$ at high field. Thus, we estimated in our previous study a background for the data, which is shown with a broken line in Fig. 3.[10] As field increases, the $S_R$ for $H$ // [110] decreases gradually and almost vanishes at $\mu_0 H \geq 0.5$ T. The small difference between the data and the background above 0.5 T may be due to the fact that our $C_m$ data was limited below $T = 20$ K. This behavior for $H$ // [110] contrasts with the other two cases for $H$ // [001] and [111]: The $S_R$ decreases more rapidly for the former, while exhibits a plateau for the latter which corresponds to the formation of the kagomé ice state.[10] It is to be noted that the peak at $T_{p2}$ in specific heat is observed just after the $S_R$ is lost, indicating that the SRO associated with $T_{p2}$ is not accompanied with a residual entropy.

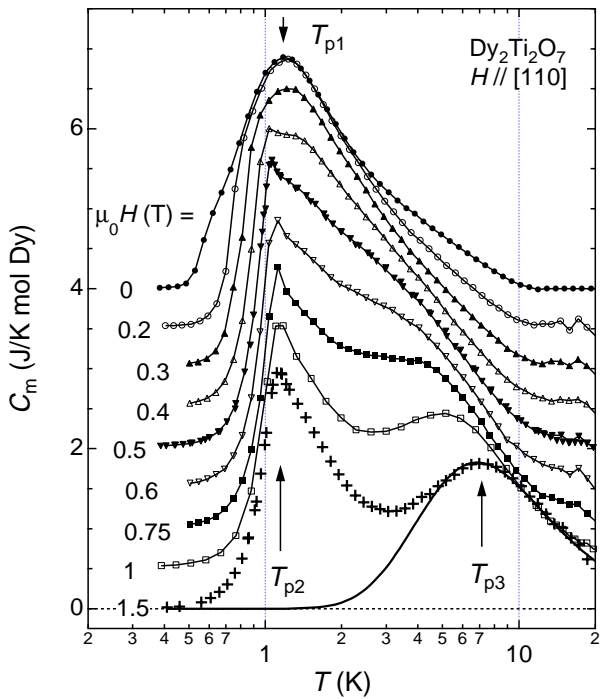

Fig. 2. Temperature dependence of magnetic specific heat $C_m$ measured at various magnetic fields of $0 \leq \mu_0 H \leq 1.5$ T in the [110] direction. Each set of data has been shifted relatively by an offset of 0.5 J/K mol Dy for clarity. Three characteristic peaks are indicated by $T_{p1}$, $T_{p2}$, and $T_{p3}$ with arrows. The solid line at the bottom shows calculated specific heat assuming a Schottky-type contribution from the $\alpha$ spin chain at $\mu_0 H = 1.5$ T.

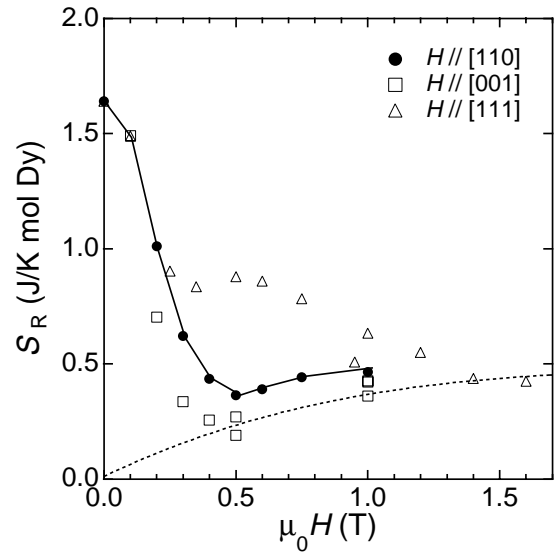

Fig. 3. Field dependence of residual entropy $S_R$ obtained for three field directions. The closed circles with lines are for $H // [110]$ from the present study, and the others for $H // [001]$ (open square) and [111] (open triangle) from the previous study.[7] The dotted line is an estimated background coming from experimental errors in determining $S_R$.

As suggested by Fennell et al.,[13] the present pyrochlore spin system under $H$ // [110] can be divided into $\alpha$ and $\beta$ chains. Thus, we assume corresponding two contributions to the specific heat; $C_m = C_\alpha + C_\beta$. The observed broad peak at $T_{p3}$ must come from the $\alpha$ chain, because it should be sensitive to the field owing to large Zeeman energy, and give a Schottky-type specific heat in the high-field limit, where interspin couplings can be ignored. This contribution has the form as;

$$C_\alpha = \frac{R(\Delta/2k_B T)^2}{2\cosh^2(\Delta/2k_B T)}, \quad (1)$$

where $R$ is the molar gas constant (8.314 JK$^{-1}$mol$^{-1}$) and $\Delta$ is the energy gap corresponding to the Zeeman splitting given by $2g_J \mu_B H$, where $g_J$ is the Landé $g$-factor with a total angular moment $J$. Note that only half mol of spins are taken into account. For a free Dy$^{3+}$ ion with $g_J J = 10$ with an angle inclined by $\theta$ (35.26°) from the field direction ([110]), one expects that $\Delta/k_B = 20\cos\theta(\mu_B/k_B)H = 10.97H$. Thus calculated $C_\alpha$ for $\mu_0 H = 1.5$ T can well reproduce the data at high temperature without any adjustable parameters, as shown in Fig. 2. Therefore, we conclude that the peak at $T_{p3}$ comes from the $\alpha$ chain.

Next, we have deduced the other contribution from the $\beta$ chain by extracting calculated $C_\alpha$ from the raw data at each field. The results are shown in Fig. 4, where three sets of data taken at $\mu_0 H = 0.75$, 1.0, and 1.5 T are shown. All the data uprise steeply with nearly the same trace at low temperature, exhibit a maximum at $T = 1.1$ K, and gradually decreases with slightly different slopes at high temperature. The disagreement at high temperature must reflect the fact that the above estimation of $C_\alpha$ is rationalized only for the high-field limit where $J$ can be ignored. In this sense the $C_\beta$ obtained for $\mu_0 H = 1.5$ T must represent the most reliable data for the true specific heat of the $\beta$ chain.

The $\beta$ spin chain is considered to be analogous to the ferromagnetic Ising spin chain, because the ice rule forces spins on the $\beta$ chain align in one direction, once an applied field makes those on the $\alpha$ chain aligned ferromagnetically: A spin on the $\beta$ chain has six neighbors interacting ferromagnetically with an identical $J/k_B$, two of which are on the same $\beta$ chain and the other four on the two nearby $\alpha$ chains, as shown in Fig.1. Thus, this spin system is in fact three dimensional. Nevertheless, the four bonds to the $\alpha$ spins is geometrically frustrated: Two are ferromagnetically and the other two are antiferromagnetically arranged. Then, it is naively expected that the $\beta$ spin chain is isolated effectively from the $\alpha$ spin chain. Actually, when we consider a domain wall in the $\beta$ chain, it violates the ice rule on one tetrahedron and gives an excitation energy of $2J$. Thus it is reasonable to treat the $\beta$ spin chain with an Ising spin chain model with the nearest-neighbor interaction, $J_1 = J$, i.e., $J_1/k_B = 1.1$ K. In this model specific heat per half mol of Dy atoms is given as

$$C = \frac{R(J_1/k_B T)^2}{2\cosh^2(J_1/k_B T)}, \quad (2)$$

which is represented in Fig. 4 with a dotted line. Apparently, the calculated curve is far from the experimental one for 1.5 T. Note that the peak value of $C$ does not change much even though the $J_1$ value is adjusted.

In order to remedy this discrepancy, we consider longer-range dipole interactions which is suggested to be important to interpret various experimental data on spin ice compounds quantitatively.[1, 2] Assuming bare dipole interactions between the $\beta$ Ising spins, $n$-th-nearest-neighbor interactions $J_n$ ($n > 1$) are obtained as $J_n/k_B = 1.41/n^3$ K for $n$ = even and $J_n/k_B = 2.35/n^3$ K for $n$ = odd. The thick line in Fig.4 shows the specific heat calculated for a model containing up to 10th neighbor interactions where the $n$ dependence of $C$ has been almost converged. The overall shape now becomes closer to the experimental curve in spite that the added higher order interactions are relatively small (for example, $J_2/k_B = 0.176$ K, $J_3/k_B = 0.087$ K, etc.). As apparent in Fig.4, however, there still remains an unignorable deviation. This suggests that a more elaborate model is necessary to reproduce the data, which may take account of the exitations of domain walls in the chain or interchain interactions, etc.

Figure 5 shows a magnetic field-temperature phase diagram, where the position of the three peaks observed in specific heat is plotted. $T_{p1}$ observed below $\mu_0 H = 0.4$ T corresponds to the original peak at zero field. $T_{p2}$ above 0.4 T is almost independent of field, while $T_{p3}$ also appearing above 0.4 T increases gradually with field. As already described, $T_{p3}$ is the freezing temperature of spins on the $\alpha$ chain. Generally, a peak for Schottky-type specific heat with an energy gap $\Delta$ is seen at $k_B T_p = 0.417\Delta$. This relation is plotted in Fig. 5, to which $T_{p3}$ approaches asymptotically at high field. In contrast, the other two field-independent peaks are related to the development of certain SRO. Similar behavior is seen in the $H$-$T$ phase diagram for $H // [111]$.[10]

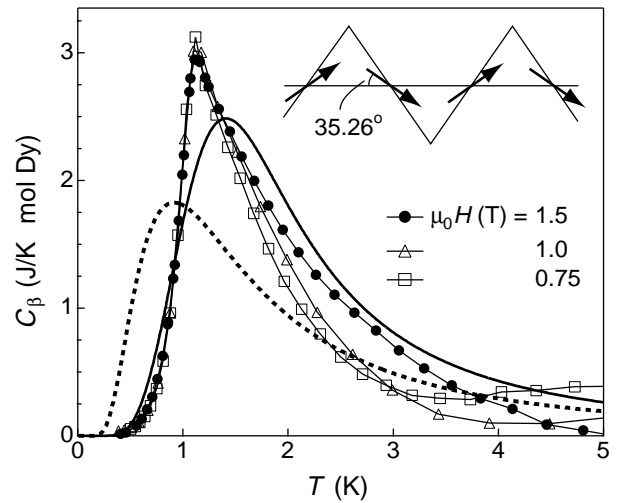

Fig. 4. Specific heat of spins on the $\beta$ chain obtained by subtracting the calculated contribution from spins on the $\alpha$ chain. Three sets of data are for $\mu_0 H = 0.75$, 1.0, and 1.5 T are shown. Note that the $C_\beta$ value is given per one mol of Dy atoms on the pyrochlore lattice, namely, per half mol on the $\beta$ chain. The dotted line represents the calculated specific heat for an Ising spin chain with only nearest-neighbor coupling $J_1/k_B = 1.1$ K, while the thick line assumes longer-range dipole interactions $J_n$ up to tenth-nearest neighbor interactions like $J_1/k_B = 1.1$ K, $J_2/k_B = 0.176$ K, $J_3/k_B = 0.087$ K, etc.

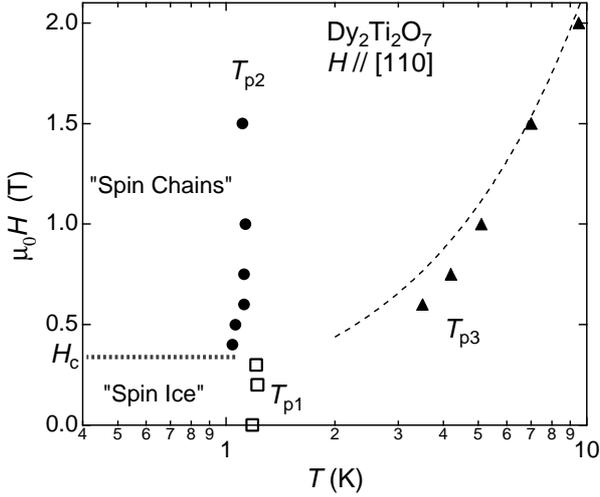

Fig. 5. Magnetic field-temperature phase diagram for $H // [110]$. Squares, circles, and triangles represent the position of three peaks found in the specific heat measurements, $T_{p1}$, $T_{p2}$, and $T_{p3}$, respectively. The dotted line shows a curve with $\mu_0 H = 0.219$ T.

Now we will consider what is actually going on with cooling in the present pyrochlore spin system under magnetic field in the [110] direction. There are two regimes in the phase diagram. Below $H_c \sim 0.4$ T the spin ice ground state is stabilized finally, because the field is not strong enough, compared with effective interaction $J$, to pin all the spins on the $\alpha$ chain along the field direction. In other words, freezing into the spin ice state occurs before complete spin alignment due to Zeeman energy when the system is cooled. However, since the partial alignment of $\alpha$ spins must take place, macroscopic degeneracy is lifted partly, giving rise to reduction in residual entropy, as observed in Fig. 3. The exact $H_c$ must be defined as the field where $S_R$ becomes zero. On the other hand, above $H_c$, complete pinning of $\alpha$ spins takes place at $T_{p3}$ on cooling. Then, only the $\beta$ spins have degree of freedom at low temperature. Below $T_{p2} \sim J/k_B$ a short-range ferromagnetic correlation develops in the $\beta$ chain under the ice rule. Consequently, one expects a unique one-dimensional ferromagnetic Ising spin system embedded in the three-dimensional pyrochlore network. The transition from the spin-ice state with residual entropy to the spin-chain state without that seems to be continuous. This is in contrast to the case of $H // [111]$, where the kagomé ice state (a reduced form from the tetrahedral spin ice state) turns into a fully polarized state (three-in, one-out/one-in, three-out state) through a first-order phase transition.[10, 15]

Neutron diffraction experiments observed Bragg scattering from the $\alpha$ chain and diffuse scattering from the $\beta$ chain under a magnetic field of 1.5 T in the [110] direction.[13] The diffuse scattering is due to the lack of coherence between the ferromagnetic $\beta$ spin chains. It was suggested that the true ground state is a $Q = X$ structure, where the net moment of the $\beta$ chain aligns antiferromagnetically with each other, though it is dynamically inhibited from being accessed on experimental timescale.[13] However, we point out here that it may be difficult for the $Q = X$ structure to become LRO, because the $\beta$ chains form a triangular lattice in the pyrochlore structure when they are viewed along the chain direction. Thus the associated frustration must destabilize antiferromagnetic couplings between chains. Therefore, we believe that the $\beta$ chain above $H_c$ behaves as a "pure" one-dimensional system without LRO in the ground state.

In conclusion, we have studied a field effect in the [110] direction on the macroscopically degenerate ground state of the spin ice in the pyrochlore oxide $Dy_2Ti_2O_7$ by means of specific heat. It is found that a one-dimensional ferromagnetic Ising spin system without LRO emerges from the spin ice ground state above $H_c \sim 0.4$ T. The specific heat for the chain deduced can be approximated theoretically by assuming long-range dipolar interactions in the chain.

**Acknowledgments**

We are grateful to T. Sakakibara for enlightening discussions. This research was supported by a Grant-in-Aid for Scientific Research on Priority Areas (A) and a Grant-in-Aid for Creative Scientific Research provided by the Ministry of Education, Culture, Sports, Science and Technology, Japan.

1) M. J. Harris, S. T. Bramwell, D. F. McMorrow, T. Zeiske and K. W. Godfrey: Phys. Rev. Lett. **79** (1997) 2554.
2) S. T. Bramwell and M. J. Gingras: Science **294** (2001) 1495.
3) A. P. Ramirez, A. Hayashi, R. J. Cava, R. Siddharthan and B. S. Shastry: Nature **399** (1999) 333.
4) M. J. Harris, S. T. Bramwell, P. C. W. Holdsworth and J. D. M. Champion: Phys. Rev. Lett. **81** (1998) 4496.
5) K. Matsuhira, Y. Hinatsu, K. Tenya and T. Sakakibara: J. Phys.: Condens. Matter **12** (2000) L649.
6) R. Siddharthan, B. S. Shastry, A. P. Ramirez, A. Hayashi, R. J. Cava and S. Rosenkranz: Phys. Rev. Lett. **83** (1999) 1854.
7) S. T. Bramwell, M. J. Harris, B. C. d. Hertog, M. J. P. Gingras, J. S. Gardner, D. F. McMorrow, A. R. Wildes, A. L. Cornelius, J. D. M. Champion, R. G. Melko and T. Fennell: Phys. Rev. Lett. **87** (2001) 047205.
8) M. Kanada, Y. Yasui, Y. Kondo, S. Iikubo, M. Ito, H. Harashina, M. Sato, H. Okumura, K. Kakurai and H. Kadowaki: J. Phys. Soc. Jpn **71** (2002) 313.
9) K. Matsuhira, Z. Hiroi, T. Tayama, S. Takagi and T. Sakakibara: J. Phys.: Condens. Matter **14** (2002) L559.
10) Z. Hiroi, K. Matsuhira, S. Takagi, T. Tayama and T. Sakakibara: J. Phys. Soc. Jpn **72** (2003) 411.
11) R. Higashinaka, H. Fukazawa and Y. Maeno: Phys. Rev. B 68 (2003) 014415.
12) M. Udagawa, M. Ogata and Z. Hiroi: J. Phys. Soc. Jpn. **71** (2002) 2365.
13) T. Fennell, O. A. Petrenko, G. Balakrishnan, S. T. Bramwell, J. D. M. Champion, B. Fak, M. J. Harris and D. M. Paul: Appl. Phys. A **74** (2002) S889.
14) H. Fukazawa, R. G. Melko, R. Higashinaka, Y. Maeno and M. J. Gingras: Phys. Rev. B **65** (2002) 054410.
15) T. Sakakibara, T. Tayama, Z. Hiroi, K. Matsuhira and S. Takagi: Phys. Rev. Lett. **90** (2003) 207205.